\documentclass[letter]{aa} 

\usepackage{graphicx}
\usepackage{txfonts}
\usepackage{color}
\begin{document} 
   \title{Discussing the first velocity dispersion profile of an ultra-diffuse galaxy in MOND}

   \author{Michal B\'ilek
          \inst{1}
          \and
          Oliver M\"uller\inst{1}
          \and
          Benoit Famaey\inst{1}
          }

   \institute{Universit\'e de Strasbourg, CNRS, Observatoire astronomique de Strasbourg (ObAS), UMR 7550, 67000 Strasbourg, France\\
              \email{bilek@astro.unistra.fr}
             }

   \date{Received tba; accepted tba}

  \abstract
{ { Using Jeans modelling, we calculate the velocity dispersion profile of the ultra-diffuse galaxy (UDG) Dragonfly~44 in MOND. For the nominal mass-to-light ratio from the literature and an isotropic profile, the agreement with the data is excellent near the center of the galaxy. However, in modified gravity, close to the cluster core, the gravitational environment should bring the galaxy back towards the Newtonian behavior. This success of the isolated MOND prediction for the central velocity dispersion could then mean that the galaxy is at large distances ($\gg5$~Mpc) from the cluster core, as hinted by the fact that close-by UDGs share similar velocities with a dispersion well below that of the cluster itself. There is however a two-sigma tension in the outer part of the UDG due to an increase of the observed dispersion profile with respect to the flat MOND prediction. This deviation could simply be a measurement error. Other possibilities could be, for a UDG far from the cluster, a higher-than-nominal baryonic mass with a tangentially anisotropic dispersion profile, or even a sign of a dark baryonic halo. If the UDG is closer to the cluster core, the deviation could be a sign that it is in the process of disruption.

   }}

   \keywords{Galaxies: individual: Dragonfly~44; Galaxies: clusters: individual: Coma cluster; Galaxies: kinematics and dynamics}

   \maketitle
%

\section{Introduction}
Ultra-diffuse galaxies (UDG) are very extended ($r_{\rm eff}>1.5$\,kpc)  low-surface brightness ($\mu_V>25$\,mag arcsec$^{-2}$) objects. They have been identified in different galactic environments for decades \citep[e.g., ][]{1984AJ.....89..919S,1988ApJ...330..634I}. {These galaxies} have recently experienced a revival \citep{2014ApJ...795L..35C,udgs,2015ApJ...807L...2K,vandenburg16,2016AJ....151...96M,2016ApJ...833..168M,2017MNRAS.468..703R,2017A&A...608A.142V,2017MNRAS.470.1512W,2018LeoI,2019MNRAS.485.1036M}. In galaxy clusters, UDGs do not contain gas while in sparser environments, they can be gas dominated \citep{udgs,shi17,leisman17, papastergis17, 2019ApJ...871L..31S}{, following the well-known density-morphology relation \citep{1980ApJ...236..351D}}.
Several formation scenarios have been proposed in the $\Lambda$CDM context: they might be tidal dwarf galaxies, galaxies formed by collapse of gas in galaxy outflows, dwarf galaxies that experienced strong tidal stripping or repeated episodes of intense star formation and others
\citep[e.g.][]{udgs,2015MNRAS.452..937Y,2016MNRAS.459L..51A,2017MNRAS.466L...1D,2018MNRAS.478..906C,2018ApJ...856L..31T}. Observations suggest that at least some UDGs form in the field and then are accreted to galaxy clusters where they experience environmental quenching \citep{2017MNRAS.468.4039R,2018MNRAS.479.4891F, alabi18}.

The dynamics of these systems provides exciting insights to the discussion. For instance, \citet{2018Natur.555..629V} announced the discovery of an apparently dark matter free UDG in the field of NGC\,1052. Soon, in the same field around NGC\,1052 a second galaxy lacking dark matter was found \citep{2019arXiv190105973V}. These two galaxies -- NGC\,1052-DF2 and NGC\,1052-DF4 -- have sparked a vast discussion in the literature \citep[e.g., ][]{2018ApJ...864L..18V,2019MNRAS.tmp..733T,2018ApJ...859L...5M,kroupa2018does,2018MNRAS.480..473F,2018arXiv181207345E,2018arXiv181207346F,2019A&A...623A..36M,2019A&A...624L...6M, 2019MNRAS.484..510N,2019MNRAS.484..245L}.

As low surface-brightness objects, UDGs are expected to show complex dynamical behavior in the context of Modified Newtonian Dynamics (MOND, \citealt{milg83a}; here we only consider MOND  modified gravity theories and not modified inertia theories). As a historic note, MOND was proposed to solve the missing mass problem in high surface brightness spiral galaxies \citep[e.g., ][]{1978ApJ...225L.107R,1981AJ.....86.1791B,1982ApJ...261..439R}  and has since excelled in reproducing the dynamics of a much larger range of galaxies (e.g., \citealt{1998ApJ...499...66M,2013ApJ...775..139M,2017ApJ...836..152L}, see also the extensive of review in \citealt{2012LRR....15...10F}). At the time MOND was proposed, the existence of UDGs was unknown. 
In MOND, a dynamical system appears Newtonian when the gravitational acceleration is greater than about the threshold of $a_0 = 1.2\times10^{-13}$\,km s$^{-2}$ \citep{milg83a,begeman91}.  Below, a departure from the Newtonian dynamics occurs such that it appears as if it hosted dark matter.
The dark matter behavior is at its full strength only if the galaxy is isolated. 
In the case the object resides in a strong gravitational field of the environment, the deviations from the Newtonian dynamics can effectively be suppressed (e.g. \citealt{milg83a,bm84}),
which is due to the non-linearity of any MOND theory \citep{milgmondlaws}.
\footnote{However, the external field effect can be weak, and even almost negligible, in some modified inertia versions of MOND \citep{modinertia}.}. 
In other words, the object will appear dark matter free, even though its internal acceleration is below $a_0$, thanks to the so-called external field effect (EFE).
To explain the dynamics of two UDGs of the NGC\,1052 group in MOND -- NGC\,1052-DF2 and NGC\,1052-DF4 (if these are not in the foreground of their putative host group, see \citealt{2019MNRAS.tmp..733T}) --, it is crucial to take the EFE into account, mostly removing the tension for NGC\,1052-DF2 {\citep{kroupa2018does,2018MNRAS.480..473F}}, and lessening it for NGC\,1052-DF4 \citep{2019A&A...623A..36M}. Interestingly, the UDG Dragonfly~44 in the Coma cluster behaves very differently from the UDGs of the NGC~1052 group{. It exhibits a pronounced deviation} from Newtonian dynamics. { Early measurements of its stellar velocity dispersion yielded $\sim 47 \,{\rm km s}^{-1}$ \citep{2016ApJ...828L...6V}, well above the nominal isolated MOND prediction. } However, all these previous studies relied on the global velocity dispersion estimated at the half light radius of the galaxy, but did not take into account the overall shape and profile of the velocity dispersion due to the difficulty in measuring it.

Now, the first velocity dispersion profile of a UDG, namely Dragonfly\,44, was presented recently by \citet{2019arXiv190404838V}. 
The data was taken with the Keck Cosmic Web Imager (KCWI) on Mauna Kea during 25.3\,hrs of observations{. T}he velocity dispersion profile was studied in the context of Newtonian gravity with several types of dark matter {profiles} in \citet{2019arXiv190404838V}. { Here we use these data to discuss Dragonfly\,44 in MOND. }

\section{Velocity dispersion profile expected in MOND in isolation}
Let us first {study the} 
velocity dispersion profile of Dragonfly\,44 {as} expected by MOND, assuming that the galaxy is far away from the Coma cluster center and any other galaxies, so that the EFE is negligible.  We calculated it using the spherically symmetric Jeans equation (e.g., \citealp{binney-tremaine08})  
\begin{equation}
\frac{1}{\rho}\frac{\mathrm{d}(\rho\sigma_r^2)}{\mathrm{d}r} + 2\,\frac{\beta(r)}{r}\sigma_r^2 ~=~ a(r),
\label{eq:jeans}
\end{equation}
where $\rho$ is the density of tracers, $\sigma_r$ the radial velocity dispersion, $\beta$ the anisotropy parameter, and $a$ the of the radial gravitational acceleration. In our case, $\rho$ is the star density of the galaxy that we approximated by a S\'ersic sphere whose parameters were obtained by fitting the photometric profile \citep{udgs,2016ApJ...828L...6V,2019arXiv190404838V}, i.e. the $I$-band luminosity of $L_I = 3.0\pm0.6\times 10^8$\,L$_\sun$, the effective radius of $R_\mathrm{e} = 4.7$\,kpc and the S\'ersic index of $n = 0.94$. We used the analytic approximation for the density of a S\'ersic sphere by \citet{sersdeproj} with the update by \citet{sersdeprojupdate}. The stellar mass-to-light ratio expected from the stellar population synthesis model of Dragonfly~44 is between 1 and 1.5 \citep{2019arXiv190404838V}. This density determines the MOND gravitational field $a(r)$ \citep{milg83a}. The Newtonian gravitational field was recalculated to the MOND field assuming the so-called simple interpolation function \citep[see, e.g.,][]{2012LRR....15...10F}. The parameter $\beta$ was considered to be zero corresponding to an isotropic profile.
 We then obtained the line-of-sight velocity dispersion using the formulas given in the Appendix  of \citet{mamon05}.

\begin{figure}
        \resizebox{\hsize}{!}{\includegraphics{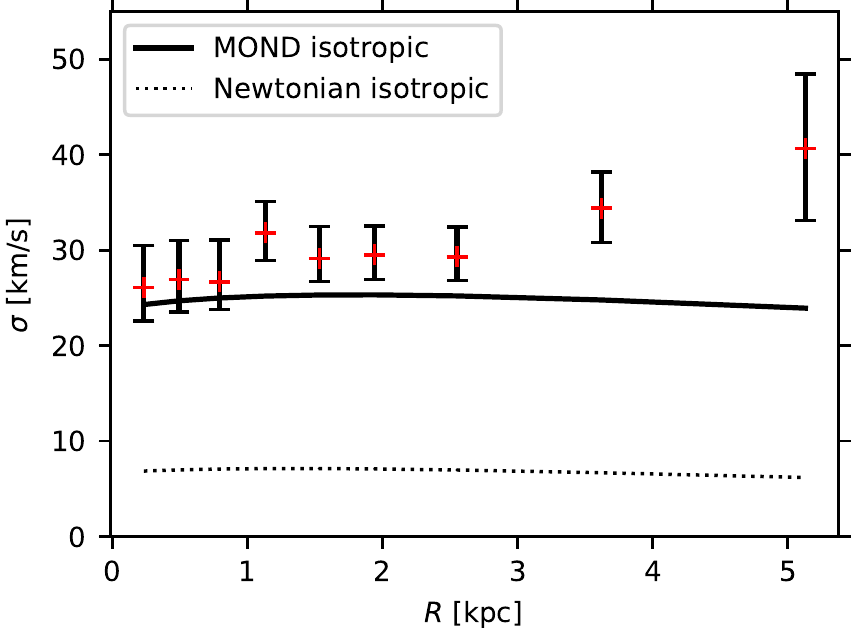}}
        \caption{Comparison of the measured velocity dispersion profile of the ultra diffuse galaxy Dragonfly~44 to the MOND no-fitting model and the Newtonian model. { A fixed mass-to-light ratio of 1.3 and the best-estimate of the galaxy luminosity as $3.0\times 10^8$\,L$_\sun$ were assumed, together with isotropy. We illustrate the effect of relaxing some of those assumptions in Fig.~2.}} 
        \label{fig:vdprof}
\end{figure}

We did not perform any fitting. Rather, we assumed a fixed mass-to-light ratio of 1.3 and the best-estimate of the galaxy luminosity as $3.0\times 10^8$\,L$_\sun$, {which lead} to our fiducial stellar mass of $M_{*,0} = 3.9\times10^{8}$\,M$_\sun$. The resulting line-of-sight velocity dispersion profile is shown in Fig.~\ref{fig:vdprof}. The agreement with observations in the center of the UDG is remarkably good given that the MOND model had no free parameters. 

{ In the dark matter framework, it is important to realize that this central velocity dispersion could have been very different from the MOND prediction, and there would actually be no obvious reason why this agreement would be so good given the unclear origin of UDGs. Indeed,} for comparison, the dotted line is an isotropic model employing Newtonian gravity without dark matter, which clearly underestimates the velocity dispersion by more than $3\sigma$. 

On the other hand, while the deviation of the MOND models from observation increases with radius, the data points are still within the $2\sigma$ uncertainty limit so that the no-fitting model is consistent with the data. { We discuss the implications of our results} and possible reasons -- { beyond the possible measurement error} -- for the deviation in the outer part of the profile in the following section.

\section{Discussion}
\label{sect:disc}
\subsection{Dragonfly\,44 far from the cluster core}
The fact that the isotropic MOND model calculated with the assumption of isolation was generally successful in Dragonfly\,44 and the Newtonian was not, suggests that the UDG has to be far {from} the Coma cluster core in the framework of MOND. If {Dragonfly\,44 was} sufficiently close to {the center of the Coma cluster} for a long time, the UDG would be subject to the EFE exerted by the  cluster so that the gravity and velocity dispersion would be lower than in isolation -- in the most extreme case they would have the Newtonian values, see, e.g.,  Fig.\,1 in \citet{2019A&A...623A..36M} for the predicted impact of the EFE onto different UDGs near massive galaxies as a function of their 3D separation. 
{To give an approximate scale for how far away Dragonfly\,44 has to be from the Coma cluster center   in order for the EFE to be negligible, we calculate the separation at which the acceleration caused by the cluster equals the acceleration in the UDG  at 1\,$R_\mathrm{e}$ in the isolated case.}
 A dedicated MOND estimate of the mass of the cluster in MOND is $4.6\times10^{14}$\,M$_\sun$  \citep{sanders03}. This  includes also the contribution of the dark matter that MOND predicts for galaxy clusters. We conclude that the distance between Dragonfly\,44 and the Coma cluster has to be well beyond 5\,Mpc. 
For comparison, the projected separation is 1.7\,Mpc (angular separation of 1$\degr$ at the distance of 100\,Mpc). 
 The wide field map of the UDGs around the Coma cluster by \citet{zaritsky19} indeed reveals that UDGs occur at least to the distance of 15\,Mpc from the cluster center. The UDGs lie in cosmic filaments of galaxies that seem to flow toward the cluster. Intriguingly, another evidence in favor of Dragonfly\,44 being quite far from the Coma cluster itself is the fact noted by \citet{2019arXiv190404838V} that the UDGs close to Dragonfly\,44 have similar radial velocities, whose differences are much smaller than the average velocity dispersion in the Coma cluster itself. If we require Dragonfly\,44 to head toward the cluster, the difference of radial velocities of the cluster and the galaxy \citep{2015ApJ...804L..26V} implies that the galaxy is in the background of the cluster.

\subsection{The effect of anisotropy}

The deviation {of the predicted vs. the observed velocity dispersion profile of Dragonfly\,44} is less than 2 sigmas at all radii, and thus it could simply be a systematic error in this difficult observation. The deviation indeed increases as the surface brightness of the object decreases, { and the data seem to imply that the dispersion profile is almost flat then suddenly start rising, which is very unlikely in the $\Lambda$CDM framework too. Nevertheless, for completeness, we now discuss in the rest of the paper the possibility that the deviation is -- at least partly -- real.}

\begin{figure}
        \resizebox{\hsize}{!}{\includegraphics{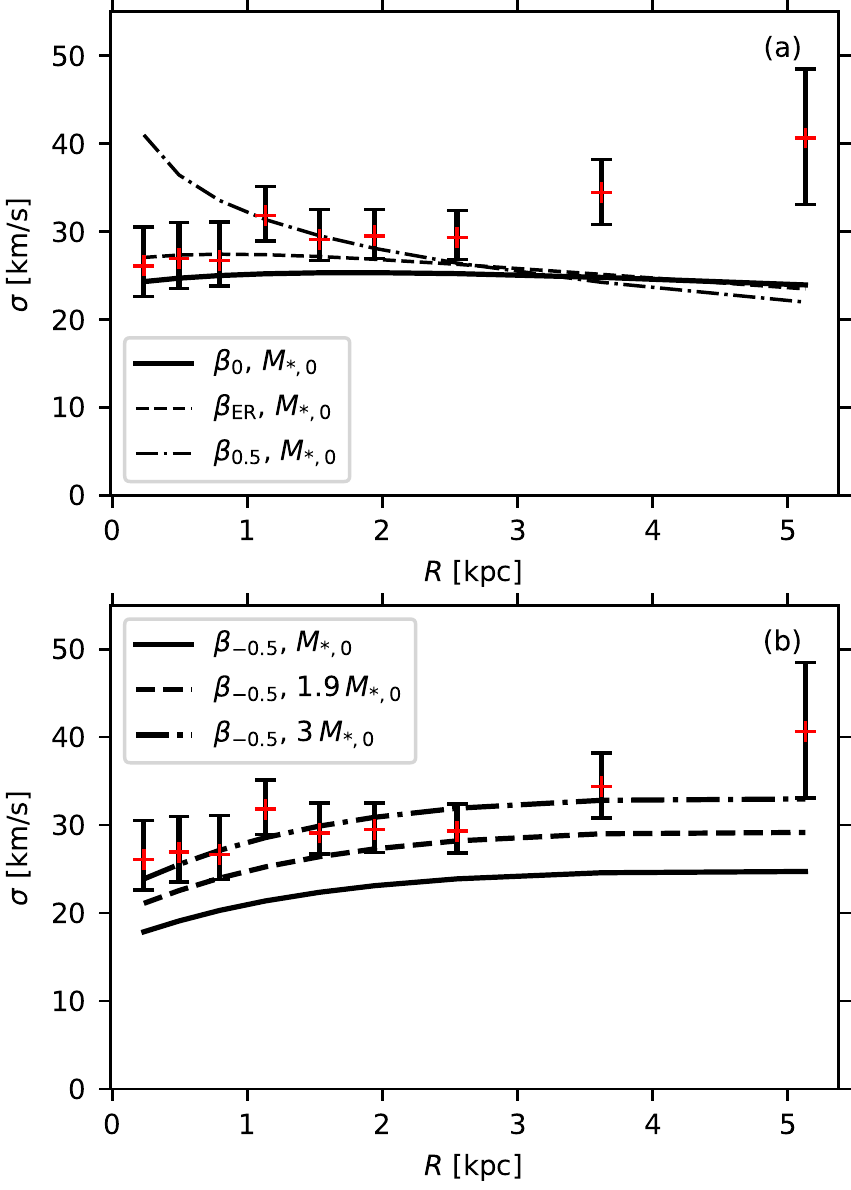}}
        \caption{Various models of the MOND velocity dispersion profile of   Dragonfly~44 discussed in Sect.~3.2. { Top panel: Radially anisotropic models. {The isotropic model -- as in Fig.\,\ref{fig:vdprof} -- is displayed as a solid line, the externally radial anisotropy model as a dashed line, and the pure radial anisotropy model as a dot-dashed line. Bottom panel: Tangentially anisotropic models. Three models for the tangential anisotropy with different stellar masses, as discussed in Sect.~3.2, are displayed with the solid, dashed, and dot-dashed lines.}}} 
        \label{fig:vdprofo}
\end{figure}

It is first mandatory to discuss the effect of the well known downside of the Jeans modeling approach, namely the ignorance of the anisotropy parameter. The effect of changing this parameter is illustrated in Fig.~\ref{fig:vdprofo}.  
When an anisotropy parameter of 0.5 is assumed, {which corresponds} to a radial anisotropy as hinted by {the} observations \citep{2019arXiv190404838V}, the {expected} velocity dispersion profile is decreasing. 
The choice of an externally radial anisotropy  $\beta_\mathrm{ER} = r/(r+1.4R_\mathrm{e})$ that is isotropic in the center and becomes radial at larger radii matches the central data points better than both the isotropic and radial anisotropies {but still deviates at larger radii}. A similar profile is expected to arise if the galaxy was in the external field dominated regime in the past \citep{wu08}. Nevertheless, it is observationally difficult to determine the anisotropy even for much brighter galaxies and as \citet{2019arXiv190404838V} note, the signs for radial anisotropy have other possible interpretations. 

The modeled velocity profile for tangential anisotropy with $\beta = -0.5$ has a shape similar to the observed profile but it has {systematically} lower values {than observed}.
{Assuming a higher M/L ratio and galaxy luminosity, each at their $2\sigma$ upper limit, leads to the stellar mass of $1.9M_{*,0}$, which  reaches the lower one sigma uncertainty limits of most of the data points.} 
Nevertheless, 
stellar M/L {ratios} are generally subjects of high uncertainties caused by {the} ignorance of the evolution of AGB stars and the initial mass function (e.g., \citealp{zibetti09, martinsson13}) such that the uncertainty in  M/L given by \citet{2019arXiv190404838V} {may be} underestimated. A higher M/L together with a larger distance caused by Dragonfly\,44 lying in the background of the cluster  makes then a stellar mass of $3M_\sun$  plausible. This value together with $\beta = -0.5$ provides an excellent model to the data, see the dash-dotted line in Fig.~\ref{fig:vdprofo}.

\subsection{Dragonfly\,44 being disrupted?}
\label{sec:efe}
{ Now we consider the possibility that Dragonfly~44 is actually closer to the cluster core. In this case, such a UDG cannot be stable, or it needs additional dark matter. Hence \citet{milgudgs} suggested that UDGs in the Coma cluster could fall on radial orbits into the Coma cluster for the first time{, where} they would then be disrupted by the combination of the EFE and tidal forces.} Again, this possibility is consistent with the facts that UDGs are observed to flow to the Coma cluster in filaments, that several UDGs close to Dragonfly\,44 share a similar radial velocity, and several other observational pieces of evidence that some UDGs form outside clusters {(see the references in the introduction)}. 
{The tidal radius \citep{2005A&A...444L..25Z} of Dragonfly\,44 is equal to $2\,R_\mathrm{e}$}
if the {three dimensional} separation from the cluster is 2.6\,Mpc. 
This {is larger}  than the projected {on-sky} separation so that the possibility that Dragonfly\,44 is being affected by tidal forces should be considered seriously.

If the velocity dispersion of the UDG was set in a region where the EFE is negligible, we do not expect that the velocity dispersion would decrease when the UDG entered in the EFE dominated regime. The fact that the gravitational well becomes shallower rather means that the object becomes unbound from its own gravity and dissolves. The outer parts of the object would obtain extra acceleration by tidal forces. This could then explain the observed rising velocity dispersion profile of Dragonfly\,44. This scenario has a few other appealing features: it would explain why Dragonfly\,44 is very elongated for a non-rotating object and why \citet{2019arXiv190404838V} noted hints that the stars of the galaxy are on radial orbits. A simulation of a satellite being disturbed by {a} field of a more massive host is depicted in Fig.\,5 of \citet{2000ApJ...541..556B},
where we can see that the object is indeed expanding and has an elongated shape.

One could suspect that the downside of this explanation is that we would have to observe the 10\,Gyr old UDG \citep{2019arXiv190404838V} in a rather special period {of} its lifetime. With the expansion velocity estimated as the deviation of the data from the equilibrium isotropic model of around 10\,km\,s$^{-1}$, the stars pass the effective radius in 460\,Myr. If {this} expansion took a much longer time the object would dissolve. Moreover, the tidal stripping cannot be very effective because the galaxy possesses many ($\sim$100) globular clusters. On the other hand, {in this scenario} we do not observe the galaxy at a random instan{ce} of its {life} but when it already entered the cluster so that signs of disruption are more expected based on the closeness of the galaxy to the cluster.  All these considerations clearly need support by future numerical simulations.

\subsection{Dragonfly\,44 with a halo of baryonic dark matter?}

The fact that many UDGs in {the Coma cluster field} lie within the projected radius where the tidal forces are expected to be substantial, might suggest that they need additional dark halos to survive in MOND (\citealp{milgudgs}, the density of UDGs as a function of cluster-centric distance was studied, e.g.,  in \citealp{vandenburg16}). 

{ Indeed, many independent pieces of evidence suggest that MOND does require dark matter in galaxy clusters that could be made of baryons (see, e.g., the review in \citealp{milgcbdm}). One possibility, suggested by \citet{milgudgs}, would be that cluster UDGs could host many of these dark baryons. } The observed rising velocity dispersion profile could then be interpreted as a consequence of the presence of such a dark halo {around the UDGs}. { However, a typical dark halo rotation curve would not reproduce the outer spike in velocity dispersions hinted at by the data.} But note that the baryonic dark halos of UDGs have likely different density profiles from what we are used for for the non-baryonic dark halos in Newtonian gravity {\citep{1997ApJ...490..493N}}.  Moreover, the fact that the velocity dispersion in the center of Dragonfly\,44 matches the MOND prediction for isolation gives us an important information: if Dragonfly\,44 is at a large 
distance from the cluster core, and that the EFE is weak, then its halo must be extended so that it has no strong gravitational influence near the center of the galaxy.  On the other hand, if the EFE acting on the UDG is substantial, then the increase of gravity due to the halo in the central parts must {\it coincidentally} compensate for the decrease of the MOND boost. 

The presence of a dark baryonic halo might be hinted to by the fact that Dragonfly\,44 has around ten times more globular clusters than dwarf galaxies of the same luminosity \citep{2018MNRAS.475.4235A}. If we interpret the empirical relation between the galaxy luminosity and number of globular clusters as a relation between the total baryonic mass of the galaxy and the number of globular clusters, then it follows from Fig.~6 of \citet{2018MNRAS.475.4235A} that the baryonic mass of Dragonfly\,44 is around $10^{10}$\,M$_\sun$, i.e. the amount of dark baryons in this galaxy is a few tens of $M_{*,0}$.

We should note that Dragonfly\,44 can be without a dark matter halo while the UDGs near the cluster center have one. Many formation scenarios for UDGs were suggested in the $\Lambda$CDM context and therefore the same should be allowed in the MOND context. At least one way to form a UDG without substantial amounts of dark baryons is known: some UDGs seem to be tidal dwarf galaxies \citep{2018ApJ...866L..11B} and these are not supposed to contain a much larger proportion of dark baryons than their parent galaxies, i.e. the spiral galaxies,  and those do not seem to contain much dark baryons (\citealp{mcgaugh00, monddarkbar}).

\subsection{How to find the answer?}

One of the main uncertainties in interpreting the data for Dragonfly\,44 is that we do not know whether the EFE exerted by the cluster is negligible,  nor whether the galaxy is in dynamical equilibrium. 

In the case of hypothetical baryonic dark matter halos suggested by \citet{milgudgs}, the easiest way to detect them would be to focus the future observations on UDGs that are in projection far enough from the cluster to be considered isolated. Then a higher than predicted velocity dispersion would be sign of a dark halo.
For the Coma cluster and a typical UDG, it means a projected separation of at least 4.5$\degr$ (assuming the median UDG mass and radius given by \citealp{udgs} of $6\times10^7$\,M$_\sun$ and $r_\mathrm{e} = 2.8$\,kpc, respectively). Figure~\ref{fig:vdprofo} suggest that the outer parts of UDGs are particularly suitable for this task since the velocity dispersion profile of Dragonfly\,44 is quite insensitive to the precise value of the anisotropy parameter (the reasons for this are probably related to the results of \citealp{2010MNRAS.406.1220W}).

Concerning UDGs in the central parts of the clusters, in order to distinguish the cases of disrupting UDGs or UDGs with dark halos, we would have to compare them to their analogs in simulations on their supposed orbits. From this perspective, it is most advantageous to observe in the future the UDGs close to cluster centers, but at distances where the internal dynamics of the UDGs is not expected to be influenced by tidal forces by the main competitors of MOND, the theories supposing Newtonian gravity and non-baryonic dark matter.

An interesting result would be if all UDGs in clusters came out to have velocity dispersions corresponding to the isolated case similarly to Dragonfly\,44. The chance that all of the central UDGs are just lying there in projection and in reality are in the outskirts is dim \citep{vandenburg16}.
Similarly, the fraction of UDGs having a dark halo whose dynamical effect would precisely compensate for the decrease of MOND gravity due to the EFE should be low. Then, two possibilities would arise: either MOND is excluded as a modification of gravity (but this would leave the question open of why the observations would agree with the isolated MOND prediction), or the UDGs are on their first radial infall and are in the process of disruption as suggested by \citet{milgudgs}. 

Note however that some theories actually expect that the internal dynamics of UDGs is not influenced by the cluster. For instance, MOND could be modified such that it does not require dark matter in galaxy clusters as proposed in \citet{ZhaoFamaey}. This could help the galaxies survive, but \citet{Hodson} showed that the high velocity dispersion, now measured at the last data point of Dragonfly\,44, could not be explained in this way, which would bring us back to the hypothesis of disruption. Another exotic possibility is that the empirical MOND effect in galaxies is due to the coupling of baryons to the phonons of a dark matter superfluid core which would encompass their baryonic profile. As extensively discussed in Sect.~IX.B of \citet{superfluid}, galaxy clusters {would} have small superfluid core{s} of about 100~kpc and no EFE would then act on the UDGs residing outside of this core, while they would still obey the isolated MOND prediction in the central parts, but have an additional effect from dark matter itself in the outskirts. Finally, the EFE can be weak in some modified inertia versions of MOND \citep[][]{modinertia}, but this would leave the question open of why UDGs in groups, such as NGC~1052, seem to be affected by the EFE, apart if they are, in fact, foreground objects.

\section{Conclusion}
Recently, the first velocity dispersion profile of an ultra-diffuse galaxy, namely Dragonfly\,44, has been measured and analyzed in the framework of Newtonian dynamics and dark matter \citep{2019arXiv190404838V}. Here we computed Jeans models for this galaxy in the framework of MOND. Assuming the galaxy to reside in isolation we found that an isotropic model reproduces the observed velocity dispersion near the center without  any {\it ad hoc} tuning of free parameters. For the outer parts, the modeled velocity dispersion deviates from the measured data points but it is still within the two sigma uncertainty limit of the measurement. While the modeled velocity dispersion profile is almost flat, the observed one is steeply rising in the outer parts. 

{ We proposed several interpretations of our results:
\begin{enumerate}
    \item the galaxy is far from the cluster (thus in isolation) and the outer deviation is simply a measurement error 
    \item the galaxy is far from the cluster, and the outer deviation is mostly an effect of tangential anisotropy for a very high stellar mass of the galaxy, which brings back the discrepancy at less than one sigma.
    \item the galaxy is closer to the cluster core, and is in the process of disruption: the inner parts are not yet observably affected by the EFE and the deviation in the outskirts is caused by tidal deformation.
    \item the deviation is a manifestation of a baryonic dark halo around the UDG, which raises the question of why it then obeys the central velocity dispersion predicted by MOND in isolation.
\end{enumerate}}
 We also discussed more exotic possibilities such as modifications of inertia without the EFE or superfluid dark matter, in which case the galaxy does not need to be far away from the cluster core to obey the isolated MOND prediction, but still requires an explanation for the outer data points mismatch. In order to test these scenarios, it is necessary to measure the velocity dispersion of more of these objects as well as to perform simulations of the evolution of UDGs moving through galaxy clusters.

\begin{acknowledgements}
O.M. is grateful to the Swiss National Science Foundation for financial support.
\end{acknowledgements}

\bibliographystyle{aa}
\bibliography{bibliographie}
  \end{document}